\documentstyle[12pt]{article}
\input epsf
\advance\textheight by \topskip
\newcounter{cureqno}
\catcode`\@=11
\newenvironment{mathletters}{\refstepcounter{equation}%
    \setcounter{cureqno}{\value{equation}}%
    \let\@curtheeqn\theequation%
    \edef\cur@eqn{\theequation}
    \def\theequation{\cur@eqn\alph{equation}}%
    \setcounter{equation}{0}}%
    {\let\theequation\@curtheeqn%
    \setcounter{equation}{\value{cureqno}}}
\catcode`\@=12
\def\lesssim{\mathrel{\hbox{\rlap{\hbox{\lower4pt\hbox{$\sim$}}}\hbox{$<$}}}}

\begin{document}
\title{Binary-pulsar tests of strong-field gravity\thanks{To
appear in: {\em Pulsar Timing, General Relativity, and the Internal Structure
of Neutron Stars}, Proceedings of a Colloquium held at the Royal Netherlands
Academy of Arts and Sciences, 24--28 September, 1996.}}
\author{Gilles Esposito-Far\`ese\\
\small\it Centre de Physique Th\'eorique\thanks{Unit\'e Propre de
Recherche 7061.}\,, CNRS Luminy\\
\small\it Case 907, F 13288 Marseille Cedex 9, France}
\date{\small(December 16, 1996)}
\maketitle

\begin{abstract}
This talk is based on my work in collaboration with Thibault Damour
since 1991. Unified theories, like superstrings, predict the existence
of scalar partners to the graviton. Such theories of gravity can be very
close to general relativity in weak-field conditions (solar-system
experiments), but can deviate significantly from it in the strong-field
regime (near compact bodies, like neutron stars). Binary pulsars are
thus the best tools available for testing these theories. This talk
presents the four main binary-pulsar experiments, and discusses the
constraints they impose on a generic class of tensor-scalar theories. It
is shown notably that they rule out some models which are strictly
indistinguishable from general relativity in the solar system. This
illustrates the qualitative difference between binary-pulsar and
solar-system tests of relativistic gravity.
\end{abstract}

\vfill
CPT-96/P.3411

gr-qc/9612039

\section{Introduction}
The usual meaning of ``testing a theory'' is rather negative: One
compares its predictions with experimental data, and a single
inconsistency suffices to rule it out. On the other hand, it is
difficult to determine what features of the theory are correct when it
passes a given test. In order to extract some positive information from
experiment, it is useful to embed the theory into a class of
alternatives. Indeed, by contrasting their predictions, it is easier to
understand in what way they differ, and to determine the common features
which make them pass or not the available tests. Moreover, this approach
can suggest new experiments to test the other features of the theories.

The best known example of such an embedding of general relativity into a
space of alternatives is the so-called Parametrized Post-Newtonian (PPN)
formalism, which is extremely useful for studying gravity in weak-field
conditions, at order $1/c^2$ with respect to the Newtonian interaction.
The original idea was formulated by Eddington \cite{edd}, who wrote the
usual Schwarzschild metric in isotropic coordinates, but introduced some
phenomenological parameters $\beta^{\rm PPN}$, $\gamma^{\rm PPN}$, in
front of the different powers of the dimensionless ratio $Gm/rc^2$~:
\begin{mathletters}
\label{eq1}
\begin{eqnarray}
-g_{00} & = & 1 - 2 {Gm\over rc^2}
+ 2 \beta^{\rm PPN} \left({Gm\over rc^2}\right)^2
+O\left({1\over c^6}\right)\ ,
\nonumber\\
&&\label{eq1a} \\
g_{ij} & = & \delta_{ij}\left[
1 + 2 \gamma^{\rm PPN}{Gm\over rc^2}+O\left({1\over c^4}\right)
\right]\ .
\label{eq1b}
\end{eqnarray}
\end{mathletters}
General relativity, which corresponds to $\beta^{\rm PPN} = \gamma^{\rm
PPN} = 1$, is thus embedded into a two-dimensional space of theories
parametrized by all real values of $\beta^{\rm PPN}$, $\gamma^{\rm
PPN}$. [The third parameter that one may introduce in front of
$2Gm/rc^2$ in $g_{00}$ can be reabsorbed in the definition of the mass
$m$.]

The constraints imposed in this space by solar-system experiments
are displayed in Fig.~\ref{fig1}, and give the following 1$\sigma$
limits on the Eddington parameters:
\begin{figure}[tb]
\begin{center}\leavevmode\epsfbox{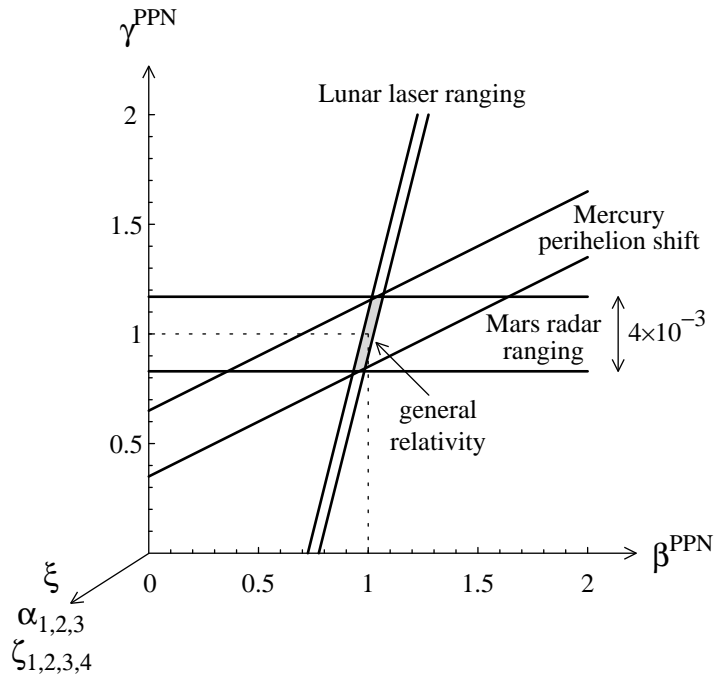}\end{center}
\caption{Solar-system constraints on the PPN parameters. The widths of
the strips have been enlarged by a factor 100. The allowed region is
shaded.}
\label{fig1}
\end{figure}
\begin{mathletters}
\label{eq2}
\begin{eqnarray}
|\beta^{\rm PPN} -1| & < & 6\times 10^{-4}\ ,
\label{eq2a} \\
|\gamma^{\rm PPN}-1| & < & 2\times 10^{-3}\ .
\label{eq2b}
\end{eqnarray}
\end{mathletters}
The PPN formalism has been further developed by Schiff, Baierlin,
Nordtvedt and Will to describe any possible relativistic theory of
gravity at order $1/c^2$. In particular, Will and Nordtvedt \cite{wn}
introduced up to 8 extra parameters (besides $\beta^{\rm PPN}$ and
$\gamma^{\rm PPN}$), each of them describing a particular violation of
the symmetries of general relativity, like local Lorentz invariance, or
the conservation of energy and momentum. Since these 8 parameters do not
have any really natural field-theoretic motivation (as opposed to
$\beta^{\rm PPN}$ and $\gamma^{\rm PPN}$; see below), we will not
consider them any longer in this paper. Let us just mention that they
are even more constrained than $(\beta^{\rm PPN}-1)$ and $(\gamma^{\rm
PPN}-1)$ by solar-system experiments. [See the contribution of J.~Bell
to the present Proceedings for a discussion of the tight bounds on some
of these parameters imposed by binary-pulsar data.] In the
10-dimensional space of all these PPN parameters ($\beta^{\rm PPN}$,
$\gamma^{\rm PPN}$, and the 8 others), only a tiny neighborhood of
Einstein's theory is thus allowed by solar-system experiments: the
intersection of the three strips of Fig.~\ref{fig1} and a very thin
8-dimensional slice parallel to the plane of this Figure. One can
therefore conclude that general relativity is essentially the only
theory which passes all these tests, and one may naturally ask the
question: Is it worth testing it any further~?

The reason why solar-system tests do not suffice is the extreme weakness
of the gravitational field in these conditions. Indeed, the largest
deviation from the flat metric is found at the surface of the Sun, and
is proportional to its gravitational binding energy $(Gm/Rc^2)_\odot
\approx 2\times 10^{-6}$ (where $R$ denotes the radius of the considered
body). In the vicinity of the Earth, the gravitational field is of order
$(Gm/Rc^2)_\oplus \approx 7\times 10^{-10}$. This explains why only the
first terms of the expansion (\ref{eq1}) are tested by solar-system
experiments. Two theories which are extremely close in weak-field
conditions can differ significantly in the strong-field regime. For
instance, the typical self-energy of a neutron star is $Gm/Rc^2\approx
0.2$, and therefore one cannot justify any more the PPN truncation of
the theory at order $1/c^2$. [A more rigorous definition of the binding
energy is $-\partial\ln m/\partial\ln G$. This expression takes its
maximum value, $0.5$, for black holes. The value $\approx 0.2$ found for
neutron stars should therefore be understood as a rather {\it large\/}
number.] Binary pulsars are thus ideal tools for testing relativistic
theories of gravity in strong-field conditions.

Before embedding general relativity into a class of contrasting
alternatives, and comparing their predictions with experimental data,
let us first describe the four main binary-pulsar tests presently
available.

\section{Binary-pulsar tests}
The aim of this talk is not to explain what is a pulsar to specialists
of the question. For our purpose, it is sufficient to note that an
isolated pulsar is essentially a (very stable) clock. A binary pulsar (a
pulsar and a companion orbiting around each other) is thus a {\it moving
clock\/}, the best tool that one could dream of to test a relativistic
theory. Indeed, the frequency of the pulses is modified by the motion of
the pulsar (Doppler effect), and one can extract from the Table Of
Arrivals many information concerning the orbit. For instance, the
orbital period $P_b$ can be obtained from the time between two maxima of
the pulse frequency. One can also measure several other Keplerian
parameters, like the eccentricity $e$ of the orbit, the angular position
$\omega$ of the periastron, etc.

\begin{figure}[tb]
\begin{center}\leavevmode\epsfxsize=250pt\epsfbox{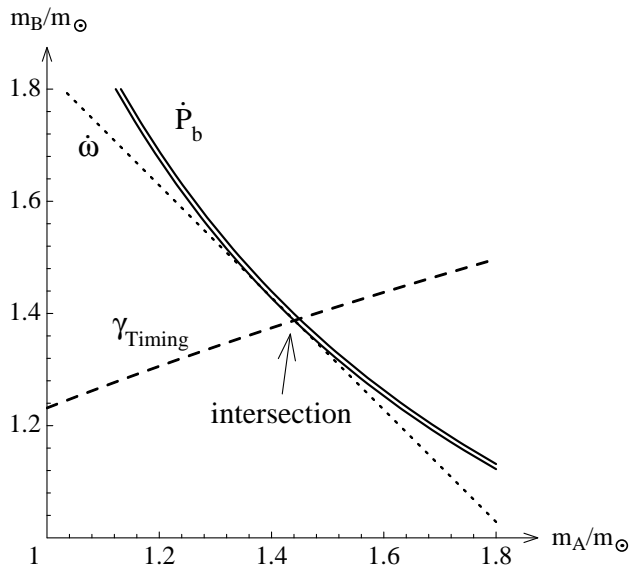}\end{center}
\caption{General relativity passes the $(\gamma_{\rm
Timing}$-$\dot\omega$-$\dot P_b)_{\rm 1913+16}$ test.}
\label{fig2}
\end{figure}

In the case of PSR B 1913+16, which has been continuously observed since
its discovery in 1974 \cite{ht}, the data are so precise that one can
even measure three relativistic effects with great accuracy. (i)~The
redshift due to the companion\footnote{$A$ denotes the pulsar, $B$ the
companion, and $r_{AB}$ the distance between them.} $\propto
Gm_B/r_{AB}c^2$ and the second-order Doppler effect $\propto v_A^2/2c^2$
are combined in an observable which has been denoted $\gamma_{\rm
Timing}$. [The index ``Timing'' is written to avoid a confusion with the
Eddington parameter $\gamma^{\rm PPN}$ introduced in Eq.~(\ref{eq1b}).]
Since the Keplerian parameters $P_b$ and $\omega$ have been measured
accurately during two decades, their time derivatives are also
available: (ii)~$\dot\omega$ gives the periastron advance (a
relativistic effect of order $v^2/c^2$), and (iii)~the variation of the
orbital period, $\dot P_b$, can be interpreted as a consequence of the
energy loss due to the emission of gravitational waves (an effect of
order $v^5/c^5$ in general relativity, but generically of order
$v^3/c^3$ in alternative theories; see below). The three
``post-Keplerian'' observables $\gamma_{\rm Timing}$, $\dot\omega$,
$\dot P_b$ can thus be compared with the predictions of a given theory,
which depend on the unknown masses $m_A$, $m_B$ of the pulsar and its
companion. However, 3 observables minus 2 unknown quantities is still 1
test. The equations $\gamma_{\rm Timing}^{\rm th}(m_A,m_B) = \gamma_{\rm
Timing}^{\rm obs}$, $\dot\omega^{\rm th}(m_A,m_B) = \dot\omega^{\rm
obs}$, $\dot P_b^{\rm th}(m_A,m_B) = \dot P_b^{\rm obs}$, define three
curves (in fact three {\it strips}) in the two-dimensional plane of the
masses $(m_A,m_B)$. If the three strips meet in a small region, there
exists a pair of masses $(m_A,m_B)$ which is consistent with all three
observables, and therefore the theory is consistent with the
binary-pulsar data. If they do not meet, the theory is ruled out.
Figure~\ref{fig2} displays these strips in the case of general
relativity, which passes the test with flying colors. [We will see below
that some other theories can also pass this test.]

\begin{figure}[tb]
\begin{center}\leavevmode\epsfxsize=250pt\epsfbox{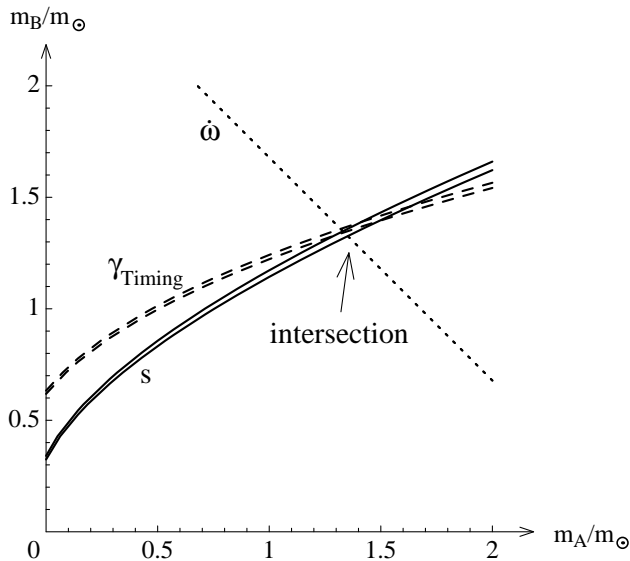}\end{center}
\caption{General relativity passes the $(\gamma_{\rm
Timing}$-$\dot\omega$-$s)_{\rm 1534+12}$ test.}
\label{fig3}
\end{figure}

The binary pulsar PSR B 1534+12 has been observed only since 1991
\cite{wol} but it is much closer to the Earth than PSR B 1913+16, and
three post-Keplerian observables have already been measured with good
precision: $\gamma_{\rm Timing}$, $\dot \omega$, and a new parameter
denoted $s$. It is involved in the shape of the Shapiro time delay (an
effect $\propto 1/c^3$ due to the propagation of light in the curved
spacetime around the companion), and it can be interpreted as the sine
$s = \sin i$ of the angle between the orbit and the plane of the sky.
[The range $r$ of this Shapiro time delay is also measured but with less
precision.] Here again, the three strips ``${\rm predictions}^{\rm
th}(m_A,m_B) =$ observed values'' can be plotted for a given theory, and
if they meet each other, the test is passed. Figure~\ref{fig3} displays
the case of general relativity, which passes the test at the 1$\sigma$
level \cite{twdw}.

As shown in Sec.~3 below, generic theories of gravity predict a large
dipolar emission of gravitational waves (of order $v^3/c^3$) when the
masses of the pulsar and its companion are very different, whereas the
prediction of general relativity starts at the much weaker quadrupolar
order ($\propto v^5/c^5$). Several dissymmetrical systems, like the
neutron star--white dwarf binary PSR B 0655+64, happen to have very
small observed values of $\dot P_b$, consistent with general relativity
but not with a typical dipolar radiation. This is the third
binary-pulsar test that we will use to constrain the space of gravity
theories.

We will also see in Sec.~3 below that in generic theories of gravity,
the acceleration of a neutron star towards the center of the Galaxy is
not the same as the acceleration of a white dwarf. This violation of the
strong equivalence principle causes a ``gravitational Stark effect'' on
the orbit of a neutron star--white dwarf system: Its periastron is
polarized towards the center of the Galaxy. [This is similar to the
effects discussed in J.~Bell's contribution to the present Proceedings.]
More precisely, the eccentricity vector ${\bf e}$ of the orbit is the
sum of a fixed vector ${\bf e}_F$ directed towards the Galaxy center
(proportional to the difference of the accelerations of the bodies), and
of a rotating vector ${\bf e}_R(t)$ corresponding to the usual
periastron advance at angular velocity $\dot \omega_R$. Several
dissymmetrical systems of this kind (such as PSRs 1855+09, 1953+29,
1800$-$27) happen to have a very small eccentricity. The only
explanation would be that the rotating vector ${\bf e}_R(t)$ is
precisely canceling the fixed contribution ${\bf e}_F$ at the time of
our observation: ${\bf e}_F+ {\bf e}_R(t) \approx {\bf 0}$. However,
this is very improbable, and one can use a statistical argument to
constrain the space of theories \cite{ds}. Moreover, by considering
several such systems, the probability that they have simultaneously a
small eccentricity is the product of the already small individual
probabilities. This idea has been used in \cite{wex} to derive a
very tight bound on the difference of the accelerations of the bodies.
This is the fourth binary-pulsar test that we will use in the following.
Of course, general relativity passes this test, since it does satisfy
the strong equivalence principle (universality of free fall of
self-gravitating objects).

These four tests are presently the most precise of all those which are
available. It should be noted that many other tests are {\it a priori\/}
possible: Damour and Taylor \cite{dt} have shown that 15 tests are in
principle possible for each binary pulsar, if the pulses are measured
precisely enough.

\section{Tensor-scalar theories of gravity}
\subsection{Introduction and action}
We saw in the previous section that several tests of gravity can be
performed in the strong-field regime, and that general relativity passes
all of them. As discussed in Sec.~1, our aim is now to embed Einstein's
theory into a class of alternatives, in order to determine what features
have been tested, and what can be further tested. A generalization of
the PPN formalism to all orders in $1/c^n$ would need an infinite number
of parameters [cf. Eq.~(\ref{eq1})]. We will instead focus on the most
natural class of alternatives to general relativity: ``tensor-scalar''
theories, in which gravity is mediated by a tensor field ($g_{\mu\nu}$)
and one or several scalar fields ($\varphi$). Here are the main reasons
why this class is privileged. (i)~Scalar partners to the graviton arise
naturally in theoretical attempts at quantizing gravity or at unifying
it with other interactions (superstrings, Kaluza-Klein). (ii)~They are
the only consistent massless field theories able to satisfy the weak
equivalence principle (universality of free fall of laboratory-size
objects). (iii)~They are the only known theories satisfying ``extended
Lorentz invariance'' \cite{n93}, {\it i.e.}, such that the physics of
subsystems, influenced by external masses, exhibit Lorentz invariance.
(iv)~They explain the key role played by $\beta^{\rm PPN}$ and
$\gamma^{\rm PPN}$ in the PPN formalism (the extra 8 parameters quoted
in the Introduction vanish identically in tensor-scalar theories).
(v)~They are general enough to describe many different deviations from
general relativity, but simple enough for their predictions to be fully
worked out \cite{def1}.

Like in general relativity, the action of matter is given by a
functional $S_m[\psi_m, \widetilde g_{\mu\nu}]$ of some matter fields
$\psi_m$ (including gauge bosons) and one second-rank symmetric
tensor\footnote{To simplify, we will consider here only theories which
satisfy exactly the weak equivalence principle, and we will restrict our
discussion to a single scalar field except in Sec.~4.} $\widetilde
g_{\mu\nu}$. The difference with general relativity lies in the kinetic
term of $\widetilde g_{\mu\nu}$. Instead of being a pure spin-2 field,
it is here a mixing of spin-2 and spin-0 excitations. More precisely, it
can be written as $\widetilde g_{\mu\nu} = \exp [2a(\varphi)]
g_{\mu\nu}$, where $a(\varphi)$ is a function of a scalar field
$\varphi$, and $g_{\mu\nu}$ is the Einstein (spin 2) metric. The action
of the theory reads thus
\begin{equation}
S = {c^3\over 16\pi G}\int d^4 x\sqrt{-g}\left(
R - 2g^{\mu\nu}\partial_\mu\varphi\partial_\nu\varphi\right)
+ S_m\left[\psi_m, e^{2a(\varphi)}g_{\mu\nu}\right]\ .
\label{eq3}
\end{equation}
[Our signature is $\scriptstyle -+++$, $R$ is the scalar curvature of
$g_{\mu\nu}$, and $g$ its determinant.]

Our discussion will now be focused on the function $a(\varphi)$, which
characterizes the coupling of matter to the scalar field. It will be
convenient to expand it around the background value $\varphi_0$ of the
scalar field ({\it i.e.}, its value far from any massive body):
\begin{equation}
a(\varphi) = \alpha_0(\varphi-\varphi_0) +{1\over 2}\beta_0
(\varphi-\varphi_0)^2 + {1\over 3!}\beta_0'(\varphi-\varphi_0)^3+\cdots\ ,
\label{eq4}
\end{equation}
where $\alpha_0$, $\beta_0$, $\beta_0'$, \dots\ are constants defining
the theory. General relativity corresponds to a vanishing function
$a(\varphi) = 0$, and Jordan-Fierz-Brans-Dicke theory to a linear
function $a(\varphi)= \alpha_0 (\varphi-\varphi_0)$. We will see in
Sec.~5 below that interesting strong-field effects occur when
$\beta_0\neq 0$, {\it i.e.}, when $a(\varphi)$ has a nonvanishing
curvature.

\subsection{Weak-field constraints}
Before studying the behavior of these theories in strong-field
conditions, it is necessary to take into account the solar-system
constraints (\ref{eq2}). A simple diagrammatic argument \cite{def3}
allows us to derive the expressions of the effective gravitational
constant between two bodies, and of the Eddington PPN parameters in
tensor-scalar theories:
\begin{mathletters}
\label{eq5}
\begin{eqnarray}
G^{\rm eff} & = & G(1+\alpha_0^2)\ ,
\label{eq5a} \\
\gamma^{\rm PPN}-1 & = & -2 \alpha_0^2/(1+\alpha_0^2)\ ,
\label{eq5b} \\
\beta^{\rm PPN}-1 & = & {1\over 2}
{\alpha_0\beta_0\alpha_0\over (1+\alpha_0^2)^2}\ .
\label{eq5c}
\end{eqnarray}
\end{mathletters}
[The factor $\alpha_0^2$ comes from the exchange of a scalar particle
between two bodies, whereas $\alpha_0\beta_0\alpha_0$ comes from a
scalar exchange between three bodies.] The bounds (\ref{eq2}) can
therefore be rewritten as
\begin{mathletters}
\label{eq6}
\begin{eqnarray}
\alpha_0^2 & < & 10^{-3}\ ,
\label{eq6a} \\
|\alpha_0^2\beta_0| & < & 1.2\times 10^{-3}\ .
\label{eq6b}
\end{eqnarray}
\end{mathletters}
The first equation tells us that the slope of the function $a(\varphi)$
cannot be too large: The scalar field is linearly {\it weakly\/} coupled
to matter. The second equation does not tell us much, since we already
know that $\alpha_0^2$ is small. In particular, it does not tell us if
$\beta_0$ is positive [$a(\varphi)$ convex] or negative [$a(\varphi)$
concave].

The same diagrammatic argument can also be used to show that any
deviation from general relativity at order $1/c^n$ ($n\geq 2$) involves
at least two factors $\alpha_0$, and has the schematic form
\begin{equation}
{\rm deviation\ from\ G.R.}= \alpha_0^2\times
\left[\lambda_0 + \lambda_1 {Gm\over Rc^2} + \lambda_2 \left({Gm\over
Rc^2}\right)^2+\cdots \right],
\label{eq7}
\end{equation}
where $Gm/Rc^2$ is the compactness of a body, and $\lambda_0$,
$\lambda_1$, \dots\ are constants built from the coefficients
$\alpha_0$, $\beta_0$, $\beta_0'$, \dots\ of expansion (\ref{eq4}).
Since $\alpha_0^2$ is known to be small, we thus expect the theory to be
close to general relativity at any order. [We do not wish to consider
models involving unnaturally large dimensionless numbers in the
expansions (\ref{eq4}) or (\ref{eq7}).] However, in two different cases
that will be discussed in Sections 4 and 5, the theory can exhibit
significant strong-field deviations from general relativity: (i)~If the
theory involves more than one scalar field, Eq. (\ref{eq6a}) does not
necessarily imply that the slope of $a(\varphi)$ is small. (ii)~Some
nonperturbative effects can develop in strong-field conditions, and the
sum of the series in the square brackets of Eq.~(\ref{eq7}) can be large
enough to compensate even a vanishingly small $\alpha_0^2$.

\subsection{Strong-field predictions}
The predictions of tensor-scalar theories in strong-field conditions
have been derived in \cite{def1}. They mimic the weak-field predictions
with the important difference that the constants $\alpha_0$, $\beta_0$
must be replaced by body-dependent parameters $\alpha_A\equiv
\partial\ln m_A/\partial\varphi_0$,
$\beta_A\equiv\partial\alpha_A/\partial\varphi_0$ (and similarly for the
companion $B$). These parameters can be interpreted essentially as the
slope and the curvature of $a(\varphi)$ at the center of body $A$ (or
body $B$). [In the weak-field regime, one has $\varphi\approx\varphi_0$,
therefore $\alpha_A\approx\alpha_0$, $\beta_A\approx\beta_0$.] In
particular, the effective gravitational constant between two
self-gravitating bodies $A$ and $B$ reads
\begin{equation}
G_{AB}^{\rm eff} = G (1+\alpha_A\alpha_B)\ ,
\label{eq8}
\end{equation}
instead of (\ref{eq5a}). The acceleration of a neutron star $A$ towards
the center $C$ of the Galaxy is thus proportional to
$(1+\alpha_A\alpha_C)$, whereas a white dwarf $B$ is accelerated with a
force $\propto (1+\alpha_B\alpha_C)$. Since $\alpha_A\neq \alpha_B$ in
general, there is a violation of the strong equivalence principle which
causes the ``gravitational Stark effect'' discussed in Sec.~2.

The strong-field analogues of $\gamma^{\rm PPN}$ and $\beta^{\rm PPN}$
are given by formulas similar to (\ref{eq5b}), (\ref{eq5c}), but
$\alpha_0^2$ is replaced by $\alpha_A\alpha_B$ and
$\alpha_0\beta_0\alpha_0$ by a combination of $\alpha_A\beta_B\alpha_A$
and $\alpha_B\beta_A\alpha_B$. The prediction for the periastron advance
$\dot \omega$ can thus be written straightforwardly.

The expression of the observable parameter $\gamma_{\rm Timing}$
involves again the body-dependent parameters $\alpha_A$, $\alpha_B$, but
also a subtle contribution proportional to $\alpha_B\times\partial\ln
I_A/\partial\varphi_0$, where $I_A$ is the inertia moment of the pulsar.
This term is due to the modification of the equilibrium configuration of
the pulsar due to the presence of its companion at a varying distance.
We have shown in \cite{def4} how to compute this effect, which happens
to be particularly large in some models (see Sec.~5 below).

The energy flux carried out by gravitational waves has been computed in
\cite{def1}. It is of the form
\begin{eqnarray}
&&{\rm Energy\ flux} = \left\{ {{\rm Quadrupole}\over c^5} +
O\left({1\over c^7}\right)\right\}_{\rm helicity\ 2}
\nonumber\\
&&+ \left\{ {{\rm Monopole}\over c} + {{\rm Dipole}\over c^3} + {{\rm
Quadrupole}\over c^5} + O\left({1\over c^7}\right)\right\}_{\rm helicity\
0}.
\label{eq9}
\end{eqnarray}
The first curly brackets contain the prediction of general relativity.
The second ones contain the extra terms predicted in tensor-scalar
theories. The powers of $1/c$ give the orders of magnitude of the
different terms. In particular, the monopolar and dipolar helicity-0
waves are generically expected to be much larger that the usual
quadrupole of general relativity. However, the scalar monopole has the
form
\begin{equation}
{{\rm Monopole}\over c} =
{G\over c}\left\{{\partial(m_A\alpha_A)\over \partial t}
+{\partial(m_B\alpha_B)\over \partial t} +
O\left({1\over c^2}\right)\right\}^2,
\label{eq10}
\end{equation}
and it reduces to order $O(1/c^5)$ if the stars $A$ and $B$ are at
equilibrium [$\partial_t(m_A\alpha_A) = 0$], which is the case for all
binary pulsars quoted in Sec.~2. [It should be noted, however, that
this monopole would be huge in the case of a collapsing star, for
instance.] The dipole has the form
\begin{equation}
{{\rm Dipole}\over c^3} =
{G\over 3c^3}\left({G_{AB}^{\rm eff}m_Am_B\over r_{AB}^2}\right)^2
(\alpha_A-\alpha_B)^2 + O\left({1\over c^5}\right)\ ,
\label{eq11}
\end{equation}
and is usually much larger that a quadrupole of order $1/c^5$ (see the
third test discussed in Sec.~2). However, when the two stars $A$ and
$B$ are very similar (e.g. two neutron stars), one has
$\alpha_A\approx\alpha_B$ and this dipolar contribution almost vanishes.
[A dipole is a vector in space; two identical stars do not define a
preferred orientation.]

\section{Tensor-multi-scalar theories}
In order to satisfy the weak-field constraints (\ref{eq6}) but still
predict significant deviations from general relativity in the
strong-field regime, the first possibility is to consider tensor-scalar
theories involving at least two scalar fields \cite{def1}. Indeed, there
can exist an exact compensation between the two fields in the solar
system, although both of them can be strongly coupled to matter. If the
kinetic terms of the scalar fields read $-(\partial_\mu\varphi_1)^2 +
(\partial_\mu\varphi_2)^2$, Eq.~(\ref{eq6a}) becomes
$|\alpha_1^2-\alpha_2^2| < 10^{-3}$, and none of the coupling constants
$\alpha_1$, $\alpha_2$ needs to be small. However, one of the fields
(here $\varphi_2$) must carry negative energy for this compensation to
occur. Therefore, these tensor-bi-scalar theories can be considered only
as {\it phenomenological\/} models, useful as contrasting alternatives
to general relativity but with no fundamental significance.

\begin{figure}[tb]
\begin{center}\leavevmode\epsfbox{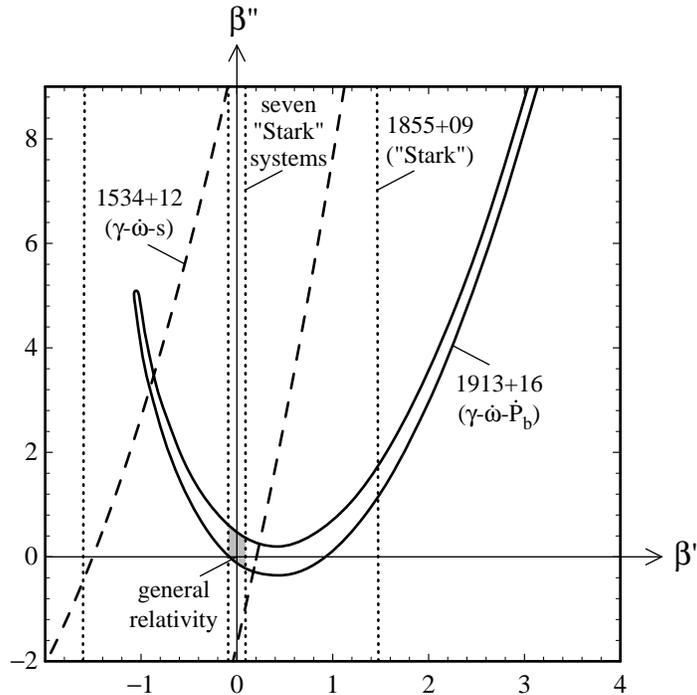}\end{center}
\caption{Constraints imposed by the binary-pulsar tests of Sec.~2
on the tensor-bi-scalar model of Sec.~4. The two dotted strips
illustrate how the precision of the ``Stark'' test is increased when
several binary pulsars are considered simultaneously. The region allowed
by all tests is the small shaded diamond around general relativity
($\beta' = \beta'' = 0$).}
\label{fig4}
\end{figure}

\begin{figure}[tb]
\begin{center}\leavevmode\epsfxsize=250pt\epsfbox{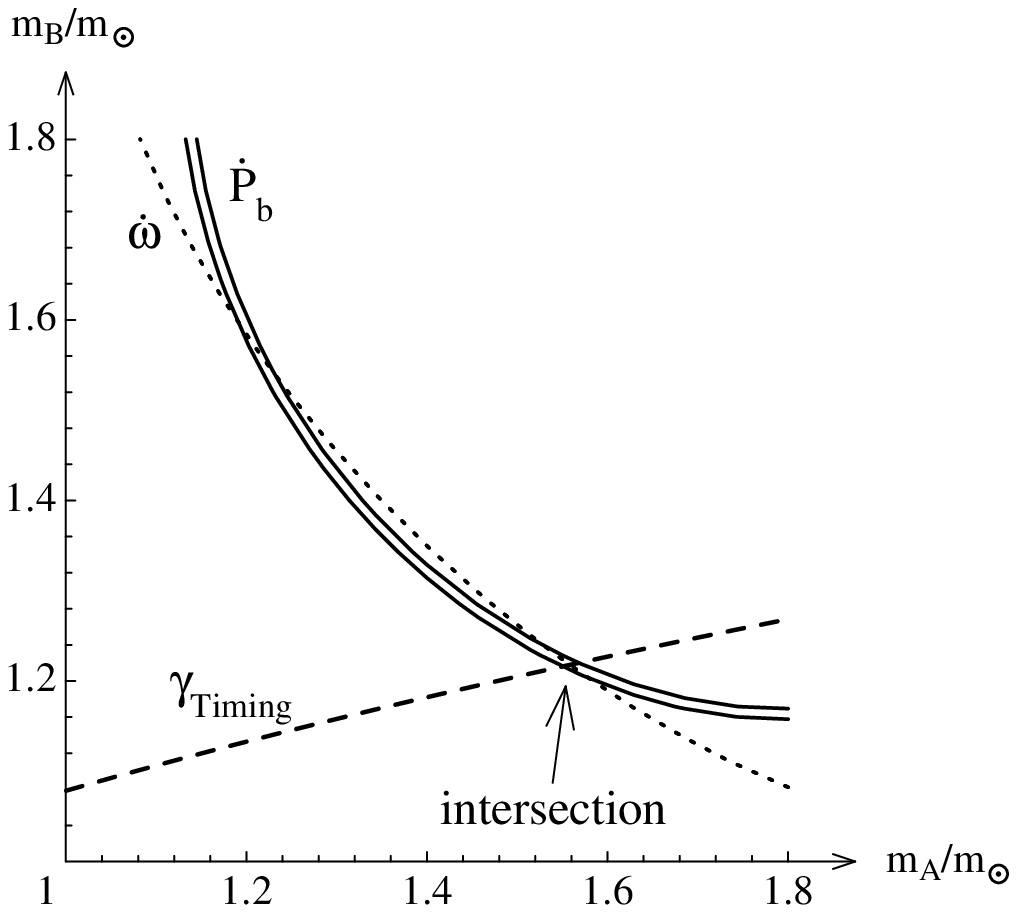}\end{center}
\caption{The tensor-bi-scalar model $\beta'=8$, $\beta''=69$, passes the
$(\gamma_{\rm Timing}$-$\dot\omega$-$\dot P_b)_{\rm 1913+16}$ test,
although the three curves are significantly different from those of
Fig.~\protect\ref{fig2}.}
\label{fig5}
\end{figure}

We have constructed in Ref.~\cite{def1} the simplest tensor-bi-scalar
model which has the following properties: (i)~It has the same
post-Newtonian limit as general relativity ($\beta^{\rm PPN}=
\gamma^{\rm PPN}= 1$), and therefore passes all solar-system tests.
(ii)~It does not predict any dipolar radiation $\propto 1/c^3$ [$\forall
A$, $\forall B$, $(\alpha_A-\alpha_B)^2=0$], and therefore passes the
third binary-pulsar test discussed in Sec.~2. Moreover, it depends on
two parameters, $\beta'$, $\beta''$, and general relativity corresponds
to $\beta'=\beta''=0$. Figure~\ref{fig4} displays the constraints
imposed by the three other binary-pulsar tests of Sec.~2 in the plane
of the parameters $(\beta',\beta'')$.
The theories passing the 1913+16
test are inside the long strip plotted in solid lines. Note that
theories which are very different from general relativity can pass this
test. For instance, Fig.~\ref{fig5} displays the mass plane $(m_A,m_B)$
for the (fine-tuned) model $\beta'=8$, $\beta''=69$.
The three strips
are significantly different from those of Fig.~\ref{fig2}, but they
still meet each other in a small region [corresponding to values of the
masses $m_A$, $m_B$ different from those found in general relativity].
To illustrate how much the theory differs from general relativity, let
us just mention that the effective gravitational constant $G^{\rm
eff}_{AB}$ between the pulsar and its companion is $1.7$ times larger
than the bare Newtonian constant $G$. We have thus exhibited a model
which deviates by 70~\% from Einstein's theory, but passes (i)~all
solar-system tests, (ii)~the ``no-dipolar-radiation'' test of PSR
0655+64, and (iii) the ``$\gamma_{\rm Timing}$-$\dot\omega$-$\dot P_b$''
test of PSR 1913+16. Before our work, this 1913+16 test was usually
considered as enough to rule out any theory but general relativity. We
have proven that other binary-pulsar tests are also necessary. In
particular, Fig.~\ref{fig4} shows clearly that the ``$\gamma_{\rm
Timing}$-$\dot\omega$-$s$'' test of PSR 1534+12 and the ``Stark'' test
complement it usefully. For instance, the model of Fig.~\ref{fig5} is
easily ruled out by the 1534+12 test: the $\gamma_{\rm Timing}$ and $s$
curves do not even meet each other (so that the observable $\dot\omega$
is not even useful here).

Thanks to the four binary-pulsar tests discussed in Sec.~2, this class
of tensor-bi-scalar models is now essentially ruled out. We have
achieved a similar result as in the weak-field regime of
Fig.~\ref{fig1}: Only a tiny neighborhood of general relativity is still
allowed. This is a much stronger result than just verifying that
Einstein's theory passes these four tests.

\section{Nonperturbative strong-field effects}
We now discuss the second way to satisfy the constraints (\ref{eq6})
while predicting significant deviations from general relativity in the
strong-field regime. As opposed to the models of the previous section,
we consider here well-behaved theories, with only positive-energy
excitations (of the type that is predicted by superstrings and
extra-dimensional theories). To simplify, we will also restrict our
discussion to the case of a single scalar field $\varphi$.

The simplest tensor-scalar theory, Jordan-Fierz-Brans-Dicke theory,
cannot give rise to nonperturbative strong-field effects for an obvious
reason. It corresponds to a linear coupling function
$a(\varphi)=\alpha_0(\varphi-\varphi_0)$, and even if the field
$\varphi_A$ at the center of body $A$ is very different from the
background $\varphi_0$, one has anyway $\alpha_A \approx a'(\varphi_A) =
\alpha_0$. Therefore, the deviations from general relativity,
proportional to $\alpha_A \alpha_B\approx \alpha_0^2$, are constrained
by the solar-system limit (\ref{eq6a}) to be $\lesssim 0.1\ \%$ even in
the vicinity of neutron stars.

On the contrary, if we consider a quadratic coupling function
$a(\varphi) = {1\over 2}\beta_0\varphi^2$, the field equation for
$\varphi$ in a body of constant density $\rho$ is of the form
$d^2(r\varphi)/dr^2 \approx \beta_0 \rho \cdot (r\varphi)$. Therefore,
the solution involves a $\sinh$ if $\beta_0>0$, and a $\sin$ if
$\beta_0<0$. More precisely, one finds
\begin{mathletters}
\label{eq12}
\begin{eqnarray}
&\alpha_A \approx \alpha_0/\cosh\sqrt{3\beta_0 Gm/Rc^2}\quad
{\rm if}\ \beta_0>0\ ,&
\label{eq12a} \\
&\alpha_A \approx \alpha_0/\cos\sqrt{3|\beta_0| Gm/Rc^2}\quad
{\rm if}\ \beta_0<0\ .&
\label{eq12b}
\end{eqnarray}
\end{mathletters}
In the case of a convex coupling function $a(\varphi)$ ({\it i.e.},
$\beta_0>0$), the deviations from general relativity are thus smaller in
strong-field conditions that in the weak-field regime: $\alpha_A\alpha_B
< \alpha_0^2<10^{-3}$. On the other hand, a concave $a(\varphi)$ can
give rise to significant deviations: If $\beta_0\lesssim -4$, the
argument of the cosine function is close to $\pi/2$ for a typical
neutron star ($Gm/Rc^2\approx 0.2$), and $\alpha_A$ can thus be large
even if $\alpha_0$ is vanishingly small. To understand intuitively what
happens when $\alpha_0 = 0$ strictly ({\it i.e.}, when the theory is
strictly equivalent to general relativity in weak-field conditions), it
is instructive to compute the energy of a typical configuration of the
scalar field, starting from a value $\varphi_A$ at the center of body
$A$ and tending towards $0$ as $1/r$ outside. One gets a result of the
form
\begin{equation}
{\rm Energy} \approx \int\left[{1\over 2}(\partial_i\varphi)^2
+\rho\, e^{\beta_0\varphi^2/2} \right]
\approx mc^2\left(
{\varphi_A^2/2\over Gm/Rc^2}
+e^{\beta_0\varphi_A^2/2}
\right)\ .
\label{eq13}
\end{equation}
When $\beta_0<0$, this is the sum of a parabola and a Gaussian, and if
the compactness $Gm/Rc^2$ is large enough, the function ${\rm
Energy}(\varphi_A)$ has the shape of a Mexican hat; the value $\varphi_A
=0$ now corresponds to a local {\it maximum\/} of the energy. It is
therefore energetically favorable for the star to create a nonvanishing
scalar field $\varphi_A$, and thereby a nonvanishing ``scalar charge''
$\alpha_A\approx \beta_0\varphi_A$. This phenomenon is analogous to the
spontaneous magnetization of ferromagnets.

\begin{figure}[tb]
\begin{center}\leavevmode\epsfxsize=300pt\epsfbox{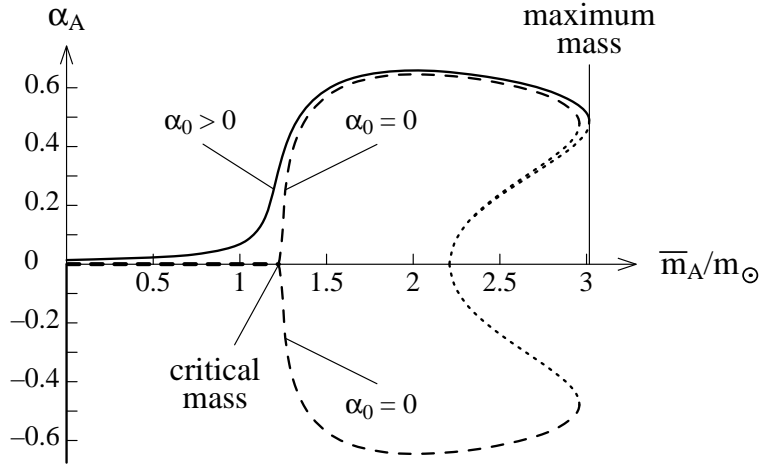}\end{center}
\caption{Scalar charge $\alpha_A$ versus baryonic mass $\overline m_A$,
for the model $a(\varphi) = -3\varphi^2$ ({\it i.e.}, $\beta_0 = -6$).
The solid line corresponds to the maximum value of $\alpha_0$ allowed by
solar-system experiments, and the dashed lines to $\alpha_0=0$
(``zero-mode''). The dotted lines correspond to unstable configurations
of the star.}
\label{fig6}
\end{figure}

We have verified the above heuristic arguments by explicit numerical
calculations \cite{def2}, taking into account the coupled differential
equations of the metric and the scalar field, and using a realistic equation
of state to describe nuclear matter inside a neutron star. We found that
there is indeed a ``spontaneous scalarization'' above a critical mass, whose
value depends on $\beta_0$. Figure~\ref{fig6} displays the scalar charge
$\alpha_A$ for the model $\beta_0 = -6$.
Note that the deviations from
general relativity are of order $\alpha_A\alpha_B\approx 35\ \%$ for a
wide range of masses from $\approx 1.25\, m_\odot$ to the maximum mass;
therefore, no fine tuning is necessary to get large deviations in a
particular binary pulsar. Note also that the nonperturbative effects do
not vanish with $\alpha_0$~: Even if the theory is {\it strictly\/}
equivalent to general relativity in the solar system, it deviates
significantly from it near compact bodies. In fact, an even more
surprising phenomenon occurs for the term $\alpha_B \partial\ln
I_A/\partial\varphi_0$ involved in the observable $\gamma_{\rm Timing}$
(see Sec.~2): This term blows up as $\alpha_0\rightarrow 0$. In other
words, a theory which is closer to general relativity in weak-field
conditions predicts larger deviations in the strong-field regime~!

\begin{figure}[tb]
\begin{center}\leavevmode\epsfxsize=250pt\epsfbox{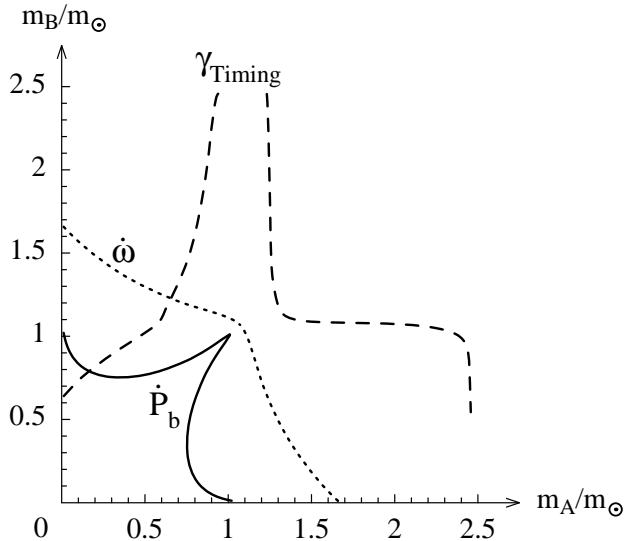}\end{center}
\caption{The model $a(\varphi) = -3\varphi^2$ does not pass the
$(\gamma_{\rm Timing}$-$\dot\omega$-$\dot P_b)_{\rm 1913+16}$ test.}
\label{fig7}
\end{figure}

The ``$\gamma_{\rm Timing}$-$\dot \omega$-$\dot P_b$'' test of PSR
1913+16 is displayed in Fig.~\ref{fig7} for the model $\beta_0=-6$ and
the maximum value of $\alpha_0$ allowed by solar-system experiments.
[The ``$\gamma_{\rm Timing}$-$\dot \omega$-$s$'' test of PSR 1534+12
gives curves similar to those of Fig.~\ref{fig7} for the first two
observables, while the $s$ strip is only slightly deviated from that of
Fig.~\ref{fig3}.] The great deformation of the $\dot P_b$ curve, as
compared to the general relativistic prediction, Fig.~\ref{fig2}, is due
to the emission of dipolar waves in tensor-scalar theories. The fact
that this dipolar radiation vanishes on the diagonal $m_A=m_B$ explains
the shape of this curve. As expected, the $\gamma_{\rm Timing}$ curve is
also very deformed because of the contribution $\alpha_B \partial\ln
I_A/\partial\varphi_0$. When $\beta_0$ is not too negative (e.g.
$\beta_0\approx -4$), a smaller value of $\alpha_0$ allows the test to
be passed: the three curves finally meet in one point. On the contrary,
when $\beta_0 <-5$, we find that the test is never passed even for a
vanishingly small $\alpha_0$ (because the term $\alpha_B \partial\ln
I_A/\partial\varphi_0$ blows up). In other words, this binary pulsar
rules out all the theories $\beta_0<-5$, $\alpha_0=0$, although they are
strictly equivalent to general relativity in weak-field conditions. This
illustrates the {\it qualitative\/} difference between binary-pulsar and
solar-system tests.

\begin{figure}[tb]
\begin{center}\leavevmode\epsfbox{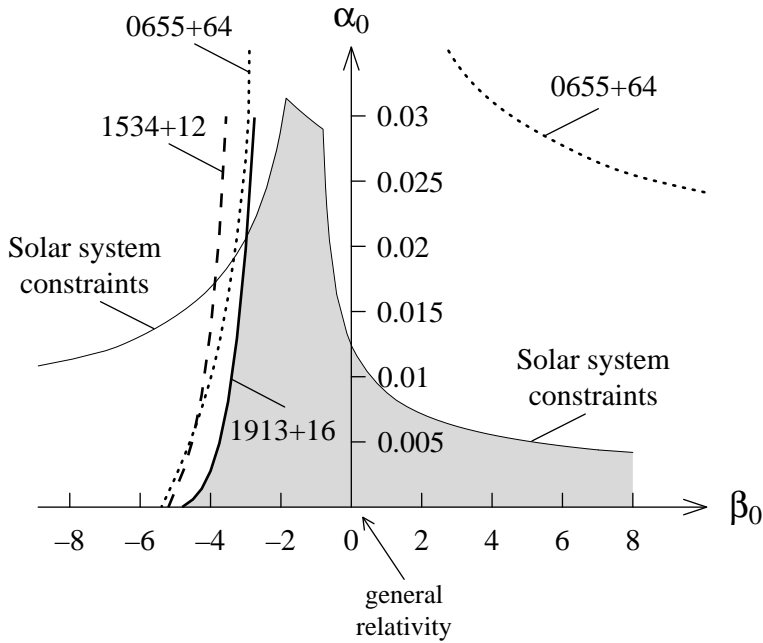}\end{center}
\caption{Constraints imposed by solar-system and binary-pulsar
experiments in the plane $(\alpha_0,\beta_0)$. In view of the reflection
symmetry $\alpha_0 \rightarrow -\alpha_0$, only the upper half plane is
plotted. The allowed regions are below and on the right of the different
curves. The shaded region is allowed by all the tests.}
\label{fig8}
\end{figure}

Generic tensor-scalar theories can be parametrized by the first two
derivatives, $\alpha_0$ and $\beta_0$, of their coupling function
$a(\varphi)$, cf. Eq.~(\ref{eq4}). It is instructive to plot the
constraints imposed by all kinds of tests in the plane
$(\alpha_0,\beta_0)$. Figure~\ref{fig8} shows that solar-system
experiments do not constrain at all the curvature $\beta_0$ of
$a(\varphi)$ if its slope $\alpha_0$ is small enough.
On the contrary,
binary pulsars impose $\beta_0>-5$, independently of $\alpha_0$. Using
Eqs.~(\ref{eq5b}), (\ref{eq5c}), this bound can be expressed in terms of
the Eddington parameters:
\begin{equation}
{\beta^{\rm PPN}-1\over \gamma^{\rm PPN}-1} < 1.3 \ .
\label{eq14}
\end{equation}
The singular ($0/0$) nature of this ratio vividly expresses why such a
conclusion could not be obtained in weak-field experiments.

Recent cosmological studies, notably \cite{dn}, have shown that theories
with a positive $\beta_0$ are easily consistent with observational data,
whereas some fine-tuning would be required if $\beta_0<0$. It is
fortunate that binary pulsars precisely privilege the positive values of
this parameter.

\section{Conclusions}
Tensor-scalar theories of gravity are the most natural alternatives to
general relativity. They are useful as contrasting alternatives, and can
suggest new experimental tests. For instance, the tensor-bi-scalar model
of Sec.~4 proved that a single binary-pulsar test does not suffice.
Well-behaved tensor-scalar theories (with no negative energy, no large
dimensionless parameters, and no fine tuning) can develop
nonperturbative strong-field effects analogous to the spontaneous
magnetization of ferromagnets. Their study illustrates the qualitative
difference between binary-pulsar and solar-system experiments: binary
pulsars have the capability of testing theories which are strictly
equivalent to general relativity in the solar system.

\end{document}